\documentclass[preprint,longabstract]{aastex}

\lefthead {Kurtz}
\righthead {Scholarly Brain}

\begin {document}

\title{The Emerging Scholarly Brain}
\author{Michael J. Kurtz}
\affil{Harvard-Smithsonian Center for Astrophysics, Cambridge, MA 02138}
\begin {abstract}

It is now a commonplace observation that human society is becoming a
coherent super-organism, and that the information infrastructure forms
its emerging brain. Perhaps, as the underlying technologies are likely
to become billions of times more powerful than those we have today, we
could say that we are now building the lizard brain for the future
organism.  \end {abstract}

\section {\label {preamble} Preamble}

\begin{verse}
{\it "... the highest level of the ant colony is the totality of its
membership rather than a particular set of superordinate individuals
who direct the activity of members at lower levels."} \\
\hfill  H\"oldobler and Wilson (1990)
\end{verse}

\section {\label {information} The Super-Organism}

 For millennia the only means of intra-species cross-generational
 communication was chemical; information was transferred via genes in
 a Darwinian process.  The structure for some very complex systems
 have been transferred by these means.

An example would be the reaction to high temperatures.  Even single
celled organisms avoid excessively hot regions; those which don't get
cooked and don't survive.  Fast forward a few eons and we have
cooperative structures of 50 trillion cells (e.g. humans) which can
very rapidly self mobilize to move a finger away from a frying pan.

A second example, indicating just how complex this chemical
communication channel can be is that primates (including humans) are
innately afraid of snakes (Hebb, 1955).  To quote Pinker (1997) "People
dread snakes without ever having seen one.  After a frightening or
painful event, people are more prudent around the cause, but they do
not fear it; there are no phobias for electrical outlets, hammers,
cars, or air-raid shelters."  Ancient creatures which had an
insufficient fear of snakes became lunch rather than becoming our
ancestors.  The agglomeration of this genetically transmitted
information has been called the ``collective unconscious'' (Jung
1935).

Creatures can learn, bees for example, can learn which color flowers
are the best sources of food, and they can communicate, as bees do
with their "waggle dance," communicating the distance and relative
position (to the sun) of a food source.  The information needed to
perform this dance appears to be entirely genetically transmitted,
there is no evidence for a learned component (von Frisch 1967).

Humans, certainly, are capable of learning communication techniques.
The development of language, aided by Darwinian processes (Pinker
1994) represents a phase transition in the information transfer
process across time which is evolution.  It became possible for
parents to tell their children "eat this mushroom, but not that one!"

This phase transition enabled humans to form large coherent groups
where information could be maintained and shared, thus problems could
be solved by distributed efforts.  Over time sophisticated bodies of
knowledge were built.
 
The next phase transition in the information transfer process was the
development of writing.  While oral transmission is powerful, and has
provided us with, as an example, the core ideas behind religion,
modern civilization is clearly too complex to be achieved without
writing.

Writing, along with the two major enabling inventions, paper and
printing, has formed the basis for the information transfer system for
our globally connected human society.  This society, viewed as an
entity onto itself, a super-organism, has capabilities far beyond
those of single individuals.  An example of these capabilities might
be the industrial revolution, which was designed and implemented
without any central direction.

We are living in another period of phase transition, one that will
happen much faster than the previous ones.  The time span for phase
transitions has been getting shorter by about a factor of 100 each
time; implying that the current one will be consummated within a single
human lifetime.

The technical underpinnings for the current phase transition are very
rapid communication (beginning, perhaps, with the telegraph) and
electronic storage.  Together our electronic communication and memory
network now form a coherent entity, a Super-Brain (Heylighten and
Bollen, 1996) for the super-organism which is human society.

Although pure communications networks have existed substantially
longer (telegraph, telephone, radio, television) it is the addition of
electronic memory which makes the computer network the fundamental
technical advance.  Within the past few decades computer networks have
coalesced to form a single entity (in graph/network studies forming
the giant component is usually viewed as a phase transition, Newman
2003).

With this the human society super-organism has developed the
functional equivalent of a brain, but we are clearly just at the
beginning of the process.  In the past 60 years the power of the
information technologies behind the transition has increased by a
factor of (about) a trillion; if this can continue for another 60
years, or indeed (Kurzweil 2001) increase more rapidly the evolving
super-organism might well be incomprehensible to us today.

The exponential growth of technical capabilities is not guaranteed,
Moore's law, for example, appears to have reached its physical limits
(in the number of transistors on a chip).  The nano-bio-cyber
convergence currently occurring suggests, however, that the growth will
continue, or, following Kurzweil (2001) accelerate.

The nature of the changes are still in the realm of science fiction;
Stapelton (1930) foresaw the fate of the "last men", after which
(evolutionarily) humans became a new race:
\begin{verse}
{\it ``The system of radiation which embraces the whole planet, and includes
the million million brains of the race, becomes the physical basis of
a racial self....

But chiefly the racial mind transcends the minds of groups and
individuals in philosophical insight into the true nature of space and
time, mind and its objects, cosmical striving and cosmical perfection....

For all the daily business of life, then, each of us is mentally a
distinct individual, though his ordinary means of communication with
others is "telepathic." But frequently he wakes up to be a
group-mind....

Of this obviously, I can tell you nothing, save that it differs from
the lowlier state more radically than the infant mind differs from the
mind of the individual adult, and that it consists of insight into
many unsuspected and previously inconceivable features of the familiar
world of men and things.'' }
\end{verse}

\section {\label{architecture} Architecture}

It appears that we humans, like the Krell (MGM 1956), will be (or
indeed are) increasingly interconnected with an electro-mechanical
system of almost unimaginable power.  Already we communicate near
telepathically with almost anyone on the planet, at almost any time,
via cell phone; already our memories are tremendously augmented by
internet search engines.

While we cannot know exactly what the future will bring, we can intuit
some things concerning the elements of our new environment, based on
past trends.  The elements we construct will be extraordinarily
complex; each with a multitude of design decisions.  This is unlike
the exact sciences, where there is a single correct description of a
physical reality; it is more like architecture, there is no single
exactly correct bridge or building design.

Certainly there are architectural elements which persist, such as a
key stone in building arches.  Perhaps the closest analog to the
information entity we are now constructing, (the super-brain?) is a 
city.  Cities are constructed by the long term and
large scale combined efforts of people.  Cities are not explicitly
designed, but they grow as a result of numerous mostly independent
efforts.  Even without explicit design cities normally have a number
of design elements in common, such as neighborhoods, or sidewalk
cafes (Alexander, et al 1977).

Major elements in all cities are places to get things.  These are
often the end points of complex webs of activity.  An example could be
a bakery, which combines efforts by fertilizer providers, seed
salesmen, farmers, millers, truckers, and others to provide bread.
Major elements in the information structure we are building are places
to get information.  These also are the end points of complex webs of
activity.  An example could be the Smithsonian/NASA Astrophysics Data
System (ADS; Kurtz, et al 1993; 2000; 2005).  The ADS combines the
efforts of lens designers, rocket scientists, astronomers, publishers
and others to provide astrophysics research information.

There are substantial differences in how these design elements are
instantiated.  There can be bakeries, hardware stores, dress shops,
toy stores, etc; and there can be Wal-Mart or Target.  Both systems
have their advantages.  In the world of information there can be
specialized providers, such as ADS, PubMed, The Internet Movie
Database, Flightstats, etc; and there can be Google or Bing.
Again both systems have advantages.

\section {\label {mind} Vectors of the Mind}

Context is an important criterion; asking to get one's tank filled
will get an altogether different result at Diver Jim's Belmont Scuba
Co than it will if asked at the gas station next door.  Likewise the
query ``plasma diagnostics'' will get a different result on PubMed
than on ADS.

Even when one can narrow down a request, for a product, or
information, to a specific realm, context is still important.  A
typical hardware store, for example, will have hundreds, if not
thousands of different fasteners (nails, screws, glue, tape, ...);
which product to purchase depends critically on exactly what the
current problem is.  Likewise the ADS has thousands of papers on a
topic such as "weak lensing", which paper to read depends critically
on the exact current problem.

Context is usually provided by people.  A customer chooses to go to a
hardware store (instead of, say, a cheese shop) to solve some problem,
which is described to a clerk.  The clerk has substantial expertise in
the specific problem domain, and suggests one, or several possible
solutions, taking into account the actual needs and abilities of the
customer.

In the information world the customer still chooses between providers,
ADS or PubMed, for example, but there is no equivalent to the clerk;
this function is performed by the information service.  Because the
actual needs and abilities of each ``customer'' vary greatly for the
exact same query (a beginning student will often be better served by
different information than would be best for a world renowned expert)
the information system will need to be able to adapt to, and perhaps
sense, the larger context of queries.

\subsection{\label{global} Global Measures}

\subsubsection{\label{eigenvector} Linear Algebra Techniques}

Clearly information systems perform functions which were previously
performed by humans, or by groups of humans.  This requires techniques
to model the thoughts and behavior of people and groups of people.
Psychologists and Sociologists have been developing these techniques
for decades.

Thurstone (1934) used eigenvector techniques to build an orthogonal
vector space model for human thought.  These techniques have become
very widely used; psychometrics and marketing for example depend
heavily on them.  They have also been used to model the idea space of
documents.  Ossorio (1966) used a psychological testing approach to
model individual opinions concerning the relevance of documents.
Kurtz (1993) reverse engineered a set of classified documents to build
a model of the thought process of a librarian classifier.  Perhaps the
most successful of these methods is latent semantic analysis (LSA;
Deerwester, et al 1990) which builds the vector space based on the
co-occurrence of words in documents.

These eigenvector techniques have also been used in many different
classification problems; such as with astronomical spectra.  Kurtz
(1982) first applied the method to stellar spectra, and Connolly, et
al (1995) rediscovered it for galaxy spectra.  While the methods have
been used, they have not found great success, compared to partitioning
line ratio diagrams (Baldwin, et al 1981) or color-magnitude diagrams
(Golay, et al 1977).  This is likely because, while the techniques are
linear, the underlying physics is highly non-linear (Kurtz 1982).

Human thought is also highly non-linear; representing it by a linear
vector space is bound to cause problems.  Although the success of LSA
as an indexing method for text demonstrated that very local measures
of nearness can be effectively defined, more generally these are not
metric spaces.  As an example, human perceptions do not follow the
Schwartz or triangle inequality.  The perceptual distance between a dog
and a vacuum cleaner is made shorter by the introduction of a third
point, a mechanical dog; a mechanical dog is a kind of dog, and a
mechanical dog is a kind of vacuum cleaner (Kurtz 1989).

\subsubsection{\label{network} Social Network Techniques}
 
By analyzing the connections between entities (people, documents,
molecules, traffic jams, ...) one can build powerful descriptive and
predictive models (Barabasi 2003, Newman 2003).  Many of the
techniques were developed for the social sciences (Wassermann and
Faust 1994) and are now widespread.  Fifteen percent of the papers
published in 2009 by Physical Review E address aspects of network
problems, as do twenty-six percent of papers published in PLOS
Computational Biology, for example.

The structure of information networks contains information about the
intelligent processes which created them.  Obtaining the underlying
structure from the network is called community detection, and is
similar to techniques for cluster analysis or classification.

Fortunato (2010) reviews these methods, currently the most popular
method is that of Girvan and Newman (2002) and the ``best'' (according
to Lancichinetti and Fortunato 2009) is that of Rosvall and Bergstrom
(RB: 2008). RB have used their algorithm on citation data to show the
interrelationships between the major fields of science; their map may
very profitably be compared with the similar map of Bolen et al (2009a)
who show a similar structure based on usage data and {\it
pre-existing} field classifications.

The RB algorithm has been used by Kurtz, et al (2007) and Henneken et
al (2009) to map the subfields of astronomy, based on both citation
data and on shared keywords for journal articles.  Attempts to build a
similar map from usage data have not, thus far, been successful,
perhaps because of the very broad readership patterns of many
astronomers.

Another measure obtainable from a network graph is the ``importance''
of the individual nodes.  Importance is normally called centrality in
this context, and there are several different centrality measures.  In
a friendship network, where people are the nodes and they are linked to
other people by friendship degree the person with the most friends
would be the person with the highest degree centrality.  Note that
friendship is directional, I may consider you my friend, but that does
not mean that you consider me a friend; thus the concepts of in-degree
and out-degree.  In a citation network the paper with the highest
in-degree is the most cited paper, while review articles would have
very high out-degree.

Betweenness centrality is another ``importance'' measure.  In a
friendship network the most central people are those with friends in
many different, otherwise autonomous cliques; in a journal to journal
citation network the most central journals are the interdisciplinary
journals (Leydesdorff 2007), like Science or Nature, which are between
otherwise autonomous fields, such as astronomy and neuroscience.
Betweenness centrality is a key measure in an information flow
network; the high betweeness centrality nodes facilitate information
transfer between fields.

The currently most used centrality measure is the expected occupation
time for each node when visited by a random walk, where the agent
randomly follows links from node to node.  This is normally called
eigenvector centrality, as the result is the same as the first
eigenvector of the node-node connectivity matrix (Bonacich 1971).
Google's famous Page-Rank algorithm (Brin and Page 1998) is essentially
eigenvector centrality, with clever implementation details.

There are several other centrality measures.  Kurtz and Bollen (2010)
give a brief introduction; Kosch\"utzki, et al (2005) a detailed
discussion.

\subsection{\label{local} Local Measures}

While measures which solve for globally optimum measures are clearly
desirable and useful, they are often not feasible.  This is
especially true in situations where the data is inherently highly
multipartite.  Journal articles, for example, may be connected by
citations, but they may also be connected by having the same author,
or having been read by the same reader, or concerning the same subject
matter, or the same astronomical object, or ....

An interesting question is how much is lost if one uses local
measures, instead of attempting a global solution.  Some recent work
discusses this.  

Eigenfactor (West, et al 2010a) is a useful measure of scholarly
journals; it is available in the Thompson-Reuters Web of Science, and
finds frequent use by librarians in making purchasing decisions.  The
Eigenfactor is essentially eigenvector centrality, measured on the
journal to journal citation graph.

Davis (2008) pointed out that the Eigenfactor journal rankings in a
subfield (medicine is what Davis used) are very similar to rankings
derived from simple citation counts.  In their two dimensional
comparison of 39 different network measures Bollen et al (2009b)
showed eigenvector centrality near to in-degree, but not identical.
West et al (2010b) showed convincingly that the Eigenfactor produces
rankings which are significantly different from plain citation counts.

While eigenvector centrality is clearly different from simple
in-degree, and in cases such as Eigenfactor likely better, the fact
remains that the differences are small.  This suggests that in many
instances very similar results may be had using much simpler measures.

The ADS second order operators (SOO: Kurtz 1992; Kurtz et al 2002)
make use of local measures in the multipartite space to achieve
specific results.  The SOO are not a ranking or clustering
methodology, per se, rather the SOO are relational operators which one
uses to build custom rankings.  Two examples of their use, both taken
from the ADS Topic Search (adsabs.harvard,edu/cgi-bin/TopicSearch),
will serve as an illustration.  Here we will use the language of
networks, Kurtz, et al (2005) use the language of lists of
attributes.  Both examples find information about a technical topic,
while this can be almost anything, for concreteness we'll use the topic
``weak lensing.''

Example 1 starts by taking the article to article citation graph,
where articles (nodes) are (directionally) connected by citations
(links).  First we take the subset of nodes which concern the topic,
weak lensing; next we sort the nodes in the subset by their in-degree
centrality, and retain only the top N (say 200).  Next we form a
super-node which is the sum of the individual top 200 nodes, and we
sort the entire graph on the out-degree centrality to the super-node.
The top of this list are articles which cite a large number of
articles on weak lensing which are, themselves, highly cited.  In
other words we have found review articles on weak lensing.

Example 2 starts with the article-reader bipartite graph, where an
article is connected to a reader if that person has read that article.
Again we initially restrict the articles to ones concerning the topic
weak lensing, then we sort that list on an attribute of the article,
but now that attribute is the publication date, we take the most
recent 200.  Again we form a super-node with the 200 articles, and
sort the reader nodes by out-degree to this super node.  We take the
readers with high out-degree to the article super-node (persons who
have read recent papers on weak lensing) and form a reader super-node.
Finally we sort all the articles in the (non restricted) graph on
in-degree from the reader super-node.  The top of this list are the
currently most popular papers among persons working in the field of
weak lensing.

Notice how these constructions solve many of the problems associated
with trust (Josang, et al 2007).  Only directly relevant, and well
regarded papers and individuals are used; ADS further restricts the
readers by removing persons who come from external search engines, who
usually do not share the goals of professional researchers (Kurtz and
Bollen 2010).

The SOO were first implemented in the ADS in 1996; they are used
throughout the system, and form much of the basis for the myADS
notification service (myads.harvard.edu).

\subsection{\label{recommend} Recommender Systems}

If direct queries to an information system, such as ADS, can be viewed
as consciously recalling data from the collective memory, then
recommender systems might be likened to having memories pop into your
head.  Recommender systems as a research and application field cover
an enormous range of different areas (c.f. recsys.acm.org) focusing
mainly on commercial uses; here we will concentrate on the recommender
systems for scholarly literature.

In a scholarly field, such as astrophysics, the information is denser,
and the use sparser, than in many other fields.  For example {\it The
Videohound's Golden Movie Retriever} lists about 30,000 movies; the
ADS contains more than four times this number of papers which contain
the word cosmology in the abstract.  The peak usage for a scholarly
article is usually the first day it becomes available (so there can be
no usage information) and decreases rapidly.  The typical ten year old
article from a major journal is downloaded once per month.

The ADS is building a set of recommenders based on the article
currently being read, in the future we will also create recommenders
based on the article viewing history of the reader.  The exact details
are yet to be determined, the systems use nearly all the techniques
described in this section, as well as the CLUTO (Zhao and Karypis
2004) clustering software.  Kurtz, et al (2009) and Henneken, et al
(2010) describe the initial implementation; which is intended for use
by active researchers.

Briefly we use the text and references in an article to find a set of
recent articles which are very similar to the target article in
subject matter.  These articles are then used with the SOO to find
recommendations using the words/keywords, citations, usage, authors,
and astronomical objects.

\section {\label{ conclusion} The Scholarly Brain}

In addition to having immensely enhanced memories, we now also have
immensely enhanced perception.  The new data intensive science (Gray
2007) rests not simply in the mechanical extension of our perception,
as begun by Galileo and van Leeuwenhoek, but on the automated
perception and analysis of huge data sets.  The ATLAS experiment at
CERN has a raw data rate of 60TB/s (Klous 2010), about fifty trillion
times the information handling capacity of humans (Fitts 1954).

The underlying technologies are being developed to satisfy the needs
of huge systems, like the LHC experiments, but ever larger systems of
sensor networks are being created to take advantage of the new
capabilities.  For these systems to combine their impact, they need to
be able to communicate with each other; they need to share a language
(Kurtz 1989; 1992).  The shared language need not be native to any
particular system or experiment (Hanisch 2001), but needs to be
universally understood, like medieval Latin, or English today.  This
standardization is taking place across all scientific and technical
fields, examples are the International Virtual Observatory Alliance's
Simple Application Messaging Protocol (SAMP; Taylor, et al 2009) or
the Open Access Initiative's Object Re-use and Exchange (OAI-ORE)
protocol (Van de Sompel, et al 2009).

Automated memory and automated perception are combining to form
automated ideas.  Standardized data descriptors along with semantic
tagging of text (Accomazzi 2009) produce a new environment, in which
inference engines, similar to the stochastic/syntactic procedures used
to analyze bubble chamber tracts (Fu and Bhargava 1973) are able to
discover new, important associations and patterns.

These systems will be able to model and predict the behavior of humans
(Barabasi 2010), indeed Ossorio (1967, 1977) spent a good fraction of
his career using his methods to model the behavior of specific humans.
The role of these systems will not be to model individual human
behaviors, but will be to act as the functional core for our
collective intelligence.

This collection of machines will not provide the totality of the brain
for the emergent super-organism which is human society; the higher
functions, the cognitive layer, will be provided by the sum of all
people.  The machines will provide the core functionality, the lizard
brain.  Analogous with our own evolution there will be innate
capabilities, like our instant reaction to heat, and anomalies, like
our fear of snakes.

Individual scholarly disciplines have long functioned as
semi-autonomous super-organisms.  The memory function (on paper) being
journal articles in university libraries.  With the advent of the
internet scholars are rapidly transforming their work habits to take
advantage of the possibilities (Borgman 2007).  New technologies
effecting the speed of information transfer (Ginsparg 1994), and the
ability to directly and effectively access huge datasets (Szalay and
Gray 2001) have already been created, more are coming daily.  The
scholarly literature has been fully digital for more than a decade.
The idea that a discipline, such as astronomy, already functions as a
coherent super-organism, with electronic memory and perception systems
is not at all far fetched.

Millions of years before he imagined it would occur we are beginning
to implement Stapledon's(1930) {\it racial mind}, beginning first not
with the human race, but with the sub races of astronomers,
bio-chemists, economists, particle physicists, etc.  With this
scholarly brain we are achieving {\it ``insight into many unsuspected
and previously inconceivable features of the familiar world of men and
things.''}

\section{\label{thanks} Acknowledgments}

This essay is dedicated to the memory of two extraordinary scientists
who showed me great kindness.  Peter Ossorio's work continues to
inspire me in several ways.  Jim Gray's (2007) $4^{th}$ Paradigm is the
clearest exposition of how the new science will actually function.

Also Andre Heck has, through the years, provided me, and many others,
with venues to discuss the deeper meaning of some current trends.  The
references to this paper alone list several.

The ADS is funded by NASA Grant NNX09AB39G.

\end {document}